\documentclass{article}
\usepackage{amsmath,amssymb}
\usepackage[dvips]{graphicx}

\usepackage{latexsym}

\begin{document}

\pagestyle{plain} 
\setcounter{page}{1}
\setlength{\textheight}{700pt}
\setlength{\topmargin}{-40pt}
\setlength{\headheight}{0pt}
\setlength{\marginparwidth}{-10pt}
\setlength{\textwidth}{20cm}

\title{Braess like Paradox in Small World  Network}
\author{Norihito Toyota   \and Hokkaido Information University, Ebetsu, Nisinopporo 59-2, Japan \and email :toyota@do-johodai.ac.jp }
\date{}
\maketitle

\begin{abstract}
 Braess  \cite{1} has been studied about a traffic flow on a diamond type network   
and found that introducing  new edges to the networks always does not achieve the efficiency. 
Some researchers studied the Braess' paradox in similar type networks by introducing various types of cost functions. 
But whether such paradox occurs or not is not scarcely studied in complex networks.  
In this article, I analytically and numerically study whether Braess like paradox occurs or not 
on Dorogovtsev-Mendes network\cite{2}, which is a sort of small world networks. 
The cost function needed to  go along an edge is postulated to be equally identified with the length between two nodes, 
independently of an amount of traffic on the edge. 
It is also assumed  the it takes a certain cost $c$ to pass through the center node in Dorogovtsev-Mendes network.  
If $c$ is small, then bypasses have the function to provide  short cuts.  
As result of numerical and theoretical analyses, while I find that any Braess' like paradox will not occur 
when the network size becomes infinite,    
I  can show that a paradoxical phenomenon appears at finite size of network. 

 \end{abstract}
\begin{flushleft}
\textbf{keywords:}
 Braess' paradox, Small world network, Dorogovtsev-Mendes network
\end{flushleft}

\section{ Introduction }
 
 When one transmits some information based on a self efficiency on some networks, 
introducing  new edges to the networks always does not achieve the efficiency. 
This feature is not restricted to information transmittance, and 
is applicable to the flows of physical objects as a traffic flow. 
This phenomenon is generally known as Braess' paradox  \cite{1} which 
has been studied about a traffic flow on a diamond type network with one diagonal line( see Fig.1). 
This is due to the fact Nash flow is necessarily not a optimal flow.    

In \cite{pas}, cases which travel times on edges are specified in such a way that the cost on the network is symmetric have been investigated. 
Their result  shows that Braess' paradox can occur in such limited cases.  
In the cases with more general cost functions where cost function on every edge is a different linear function every adge, 
the conditions that Braess' paradox occurs have been closely investigated in Braess network configuration as Fig.1\cite{Zrer}. 
Moreover Valiant and Royghtgarden \cite{Vali}  have proved that Braess' paradox is likely to occur in  a natural random network. 
An instructive review and many references are given in \cite{Bloy} 

A study that can be interpreted as a phenomenon like Braess' paradox has been done \cite{2}. 
They analytically investigated when the average shortest path is optimal on Dorogovtsev-Mendes network\cite{2}, 
which is a kind of small world networks.
Moreover the authors in \cite{3,4} analytically and numerically studied the situation where some cost is required  
when one goes via the center of the network. 
They pointed out that increasing the bypass via the center does not reduce the average cost. 

The researches so far are with the proviso, however, that  information or some objects on a network returns to the start node. 
In this article, I analytically and numerically study  the cases that a start point(node) of information/objects is different from a goal node 
on Dorogovtsev-Mendes network as the original Braess' paradox.
Through it, I show that though any Braess' like paradox does not occur when the network size becomes infinite,  
a paradoxical phenomenon appears at finite size of the network. 
\begin{figure}[ht]
\centering
\includegraphics[scale=1]{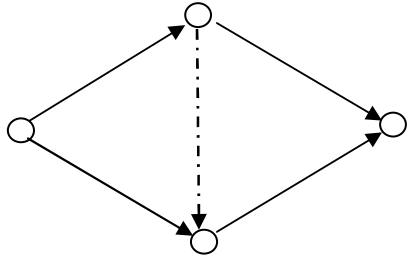}
\caption{Braess' Network Configuration}
\end{figure}

\section{Model based on Dorogovtsev-Mendes networks}
%
So far the diamond type network where a dashed line represents a bypass as shown in Fig.1 has been mainly investigated.  
Dorogovtsev-Mendes network\cite{2} is a small world like network introduced by Watts and Strogatz\cite{Watt1,Watt2} 
but with one center node. 
Some edges are drawn from circumferential nodes to the center with probability $p$ like Fig.1.  
The edges on the circumferential nodes are directed but the edges drawn from circumferential nodes to the center 
are not so.  
So while information/objects is/are allowed to move in only one direction on the circumferential circle,  
one allows traffic flow in both direction on the edges drawn from circumferential nodes to the center.  
The all lengths between adjacent circumferential nodes are one and the all ones between circumferential nodes and the center node are 0.5. 
We consider the cost needed to go from S to T to discuss Braess like paradox. 
I postulate that the cost needed to go along an edge  is just equivalent to the length between two nodes, 
independently of an amount of traffic on every edge. 
For that reason, we do not need to consider the total amount of traffic that comes into S or the network.  
It is also assumed  the cost to pass through the center node is $c$. 
If $c$ is small, then  bypasses come to provide short cuts.  

\begin{figure}[ht]
\centering
\includegraphics[scale=0.5]{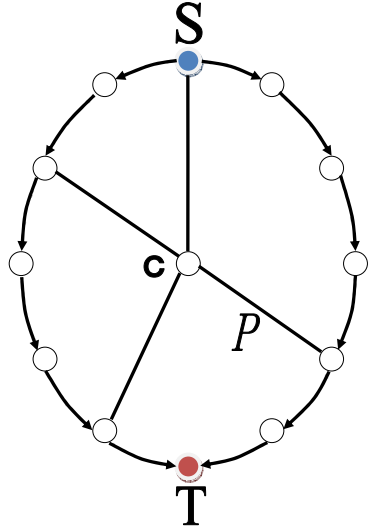}
\caption{Model based on Dorogovtsev-Mendes networks}
\end{figure}
\section{Average cost}
%
There are $n$ circumferential nodes and one center node in Dorogovtsev-Mendes network. 
In this network, the possibility $P(\ell)$ that takes the length $\ell$ to go from S to T is estimated by the following expression;
\begin{align}
P( \ell ) = & (1+c)p^2\delta_{\ell,1}+2np(1-p)^{2n-1}    \delta_{\ell,n} + (1-p)^{2n}  \delta_{\ell,n} 
+  2p^2(2-p)(1-p)^{2\ell-3}    \Theta(\ell-2, n+1-\ell)  \notag \\
 +&\Theta(\ell-3, n+1-\ell)  \times p^2  ( \ell-2 ) (p-2)^2(1-p)^{2\ell -4},    
\end{align}
wher $\Theta (a,b)\equiv \theta(a)-\theta(b)$ with $a<b$ and $\theta(\ell-x)$ is the Heaviside function defined by 
\begin{equation}
\theta(\ell-a) =\begin{cases}
   1& \ell \geq x \\
   0& \ell < x. 
        \end{cases} 
\end{equation}
\begin{figure}[b]
\centering
\includegraphics[scale=0.7]{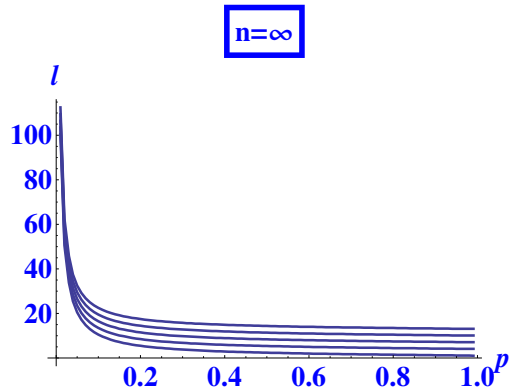}
\caption{  Behaviors of $\langle \ell_{p,c} \rangle$   at $n\rightarrow\infty$ （ $c=0,3,6,9,12$ from above）}
\end{figure}
As $\ell \geq3$ for simplicity, the average shortest path length $\langle \ell \rangle$ between S and T is given by 
\begin{align}
\langle \ell_{p,c} \rangle &= \sum_{\ell=3}^{n}  \ell P( \ell )=\Bigl( \frac{1+2q-q^2}{1-q^2} +c\Bigr)-\frac{(q^2+1)q^{2a-2}}{1-q^2}   -\frac{2q^{2a-1}}{(1+q)} 
 -(a-2)(2a-n-4)q^{2a}\notag \\
 +& \bigl(  (a-2)(2a-n-4)-(n-2+a)-n(a-2) \bigr) q^{2a-2}  +n(a-1)q^{2a-4}  + n(1-n)q^{2n-4}   \notag \\
+&n\Bigl( (n-2)  + (1-q)\bigl(2n-(n+1)(1-q) \bigr) \Bigr) q^{2n-2}+nq^{2n}+ -2(1-q)nq^{2n-1},  
\end{align}
where $a=n-c$ and $q=(1-p)$. The first term in the right side of (3) is independent of the network size $n$. 

At the limitation of $n\rightarrow \infty$, we obtain  
\begin{align}
\langle \ell_{p,c} \rangle  &
=\frac{1+2q-q^2}{1-q^2}+c. 
\end{align}
To find an extremum of $\langle \ell_{p,c} \rangle$, we  differentiate it  with respect to $p$; 
\begin{equation}
\frac{d \langle \ell_{p,c} \rangle _{\infty} }{dp} =-\frac{d \langle \ell_{p,c} \rangle _{\infty} }{dq}=-\frac{2(1+q^2)}{(1-q^2)^2}-\frac{dc}{dq}. 
\end{equation}
The derivative is negative when $c$ is a constant and  $\langle \ell_{p,c} \rangle$ decreases monotonically as $p$.
Fig. 3 shows the results of numerical calculations. 
In this case, more bypasses increases, more information/objects propagate(s) well.
 
When $c$ depends on $p$ so $q$,  the equation to find an extreme value of $\langle \ell_{p,c} \rangle$ is 
\begin{equation}
\frac{dc}{dq}=-\frac{2(1+q^2)}{(1-q^2)^2}. 
\end{equation}
Solving the differential equation, we obtain
\begin{equation}
c= \gamma  -\frac{2q}{1-q^2},
\end{equation}
where $\gamma $ is an integral constant.  
Substituting the $c$ into (4), we get 
\begin{equation}
\langle \ell_{p,c} \rangle =\gamma  +1>0. 
\end{equation}
The last inequality is due to a natural condition $c>0$ in (7).\\

Next we consider the case of $n=c+a \geq c$ where $n$ takes a finite value but $a$ takes some constants. 
Differentiate $\langle \ell_{p,c} \rangle$ with respect to $q$ to find extreme values, we obtain
\begin{align}
\frac{d \langle \ell_{p,c} \rangle}{dq} &= \frac{2}{(q^2-1)} \times \Bigl(  
1+q^2+(a-2)(a-1)nq^{2a-5}+(a-1)\bigl( 9+2a^2+7n-a(4n+9) \bigr) q^{2a-3} \notag \\
+& (1-2a)q^{2a-2}+  \bigl( 18-6a^3+8n+6a^2(n+5)-a(15n+44) \bigr)q^{2a-1} +(2a-3)q^{2a} \notag \\
+&  \bigl( 6a^3 -3(n+3) -a^2(4n+27) +a(9n+34)\bigr) q^{2a+1}  -a(a-2)(2a-n-4)q^{2a+3} \notag \\
-&  n(n-1)(n-2)q^{2n-5}+n(4n-7)(n-1)q^{2n-3}-n(6n^2-15n+8)q^{2n-1}  \notag \\
 + &n(4n^2-9n+3)q^{2n+1}-n^2(n-2)q^{2n+3}
\Bigr) \\
=& \sum_{k=1,3,\cdots}^{2a-3}kq^{k-1}+2 \Bigl( - n(n-1)(n-2)q^{2n-5}+n(2n-3)(n-1)q^{2n-3}-n^2(n-2)q^{2n-1} 
\Bigr). 
\end{align}


Fig.4 shows $p$ vs. $\langle \ell_{p,c} \rangle$, which represents the average of the total cost from S to T, graph for diverse $a$. 
From Fig.4,  we can observe that $\langle \ell_{p,c} \rangle $ reaches the maximum value at $p \simeq 0.1\sim0.2$, as $a$ is small. 
Thus increasing bypasses does not always enhance traffic efficiency of the network and it costs more with  excessive edges. 
 We can interpret that Braess like paradox occurs in that meaning. 
Though $n$  also takes a larger value for larger $a$ by its very nature, we find that the value of $p$ at extrema approaches zero 
in progression  from Fig.4. 
Fig.5 shows  $p$ at the peaks in Fig.4 that are given by solving (10)$=0$ for various $n$ when $a=4$.   
The result corresponds with those of Fig.4, well.
The Fig.4 shows that $p$ at extrema converses to zero as $a$ becomes larger. 
 So the paradox disappears. 
This is corresponds to the fact that there is no paradoxical behavior at $n \rightarrow \infty$.  

\begin{figure}[h]
\centering
\includegraphics[scale=0.67]{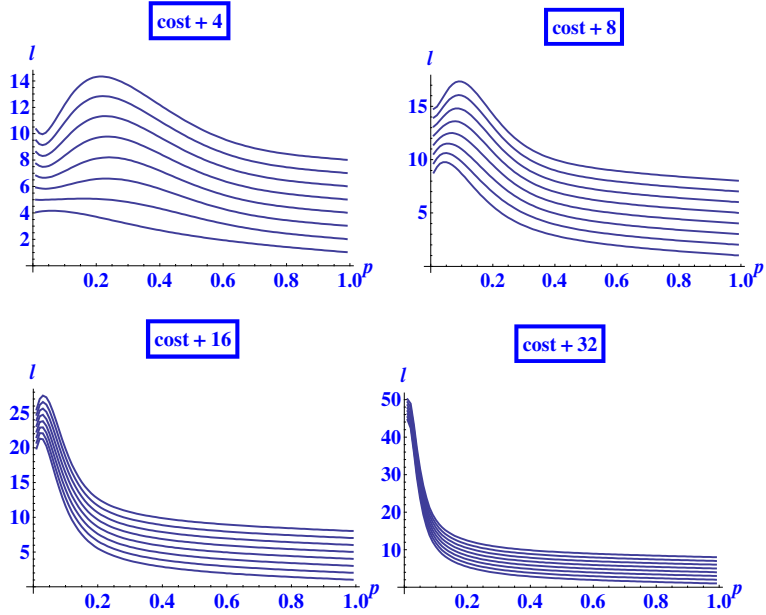}
\caption{ Behaviors of $\langle \ell_{p,c} \rangle$ at $n=cost+(4,8,16,32)$ （ from above $c=0,1,2,\cdots,7$ at each graph）}
\end{figure}

\begin{figure}[h]
\centering
\includegraphics[scale=0.8]{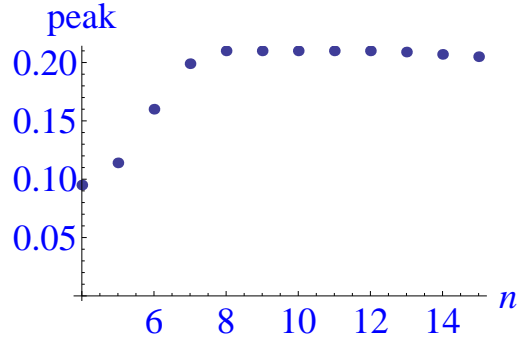}
\caption{  $p$ values  at peaks for various $n$ }
\end{figure}

\section{Summaries}

 Braess' paradox  \cite{1}, which introducing  new edges (one diagonal line) to the networks always does not achieve the efficiency, 
has been originally studied about a traffic flow on a diamond type network.  
Some researchers studied Braess' paradox in the similar type networks by introducing various types of cost functions. 
But whether such paradox occurs or not was not scarcely studied in complex networks. 

In this article, I analytically and numerically studied the cases that a start point(node) of information/objects is   
different from a final node on Dorogovtsev-Mendes network that is a sort of small world networks as the original Braess' paradox.
Here the cost function needed to go along an edge is postulated to be equally identified with the length between two nodes, 
independently of an amount of traffic on the edge. 
So, we do not need to consider the total amount of traffic that comes into S or the network.  
It is also assumed  the cost to pass through the center node is $c$. 
If $c$ is small, then bypasses have the function to provide short cuts.  
We explored whether Braess like paradox occurs or not in Dorogovtsev-Mendes network under the situation. 
As result, I showed that any Braess like paradox will not occur at the large network size limit.   
I could, however,  also show that a paradoxical phenomenon appears at finite size of network.  

The studies of more general situations,  especially such a situation as the costs to go along edges depend on amounts of traffic on the edges, 
would give more a great  wealth of knowledge for Braess' paradox. 
%
%

%
\end{document}